\documentclass[prl,superscriptaddress,twocolumn,footinbib]{revtex4-1}
\pdfoutput=1
\usepackage[colorlinks=true,urlcolor=blue]{hyperref}
\usepackage{amsfonts}
\usepackage{graphicx}
\usepackage{placeins}
\usepackage{amsmath}
\usepackage{xcolor}
\usepackage{bm}
\usepackage{url}

\usepackage{color}
\definecolor{red}{rgb}{0.75,0,0}
\definecolor{blue}{rgb}{0,0,0.75}
\definecolor{green}{rgb}{0,0.5,0}

\begin{document}

\title{Geometrical control of active turbulence in curved topographies}

\author{D. J. G. Pearce}
\affiliation{Instituut-Lorentz, Universiteit Leiden, P.O. Box 9506, 2300 RA Leiden, The Netherlands}
\author{Perry W. Ellis}
\affiliation{School of Physics, Georgia Institute of Technology, Atlanta, Georgia 30332, USA}
\author{Alberto Fernandez-Nieves}
\affiliation{School of Physics, Georgia Institute of Technology, Atlanta, Georgia 30332, USA}
\author{L. Giomi}
\affiliation{Instituut-Lorentz, Universiteit Leiden, P.O. Box 9506, 2300 RA Leiden, The Netherlands}

\begin{abstract}
We investigate the turbulent dynamics of a two-dimensional active nematic liquid crystal constrained on a curved surface. Using a combination of hydrodynamic and particle-based simulations, we demonstrate that the fundamental structural features of the fluid, such as the topological charge density, the defect number density, the nematic order parameter and defect creation and annihilation rates, are approximately linear functions of the substrate Gaussian curvature, which then acts as a control parameter for the chaotic flow. Our theoretical predictions are then compared with experiments on microtubule-kinesin suspensions confined on toroidal active droplets, finding excellent qualitative agreement.
\end{abstract}

\maketitle

Experimental studies on active liquid crystals \cite{Surrey:2001,Sanchez:2012,Sumino:2012,Keber:2014,Ellis:2018,DeCamp:2015,Guillamat:2016,Guillamat:2017,Schaller:2010,Schaller:2013,Dombrowski:2004,Wogelmuth:2008,Zhang:2010,Wensink:2012,Dunkel:2013,Wioland:2016,Riedel:2005,Creppy:2015,Zhou:2014} have unveiled, in the past decade, a cornucopia of novel hydrodynamic phenomena with no counterparts in passive complex fluids. Active liquid crystals are orientationally ordered fluids consisting of self- or mutually-propelled rod-shaped constituents, generally of biological origin. Examples include {\em in vitro} mixtures of microtubules and kinesin \cite{Surrey:2001,Sanchez:2012,Sumino:2012,Keber:2014,DeCamp:2015,Guillamat:2016,Guillamat:2017}, actomyosin gels \cite{Schaller:2010,Schaller:2013}, suspensions of motile cells, such as flagellated bacteria \cite{Dombrowski:2004,Wogelmuth:2008,Zhang:2010,Wensink:2012,Dunkel:2013,Wioland:2016} and sperm \cite{Riedel:2005,Creppy:2015}, and ``living liquid crystals'' obtained from the combination of biocompatible chromonic liquid crystals and bacteria \cite{Zhou:2014}. Depending on the abundance of biochemical fuel as well as the system density and geometry, these active liquids have been observed to self-organize into an extraordinary variety of spatiotemporal patterns, such as traveling bands \cite{Schaller:2010} and vortices \cite{Wioland:2016}, oscillating textures \cite{Schaller:2010,Keber:2014}, ordered arrangements of topological defects \cite{Riedel:2005,DeCamp:2015} and turbulent flows at low Reynolds number \cite{Sanchez:2012,Dombrowski:2004,Wogelmuth:2008,Wensink:2012,Dunkel:2013,Creppy:2015,Zhou:2014}.

Starting from the pioneer work by Keber {\em at al.} \cite{Keber:2014} on active nematic vesicles, unraveling the interplay between substrate geometry and the collective motion inherent to active fluids has surged as one of the fundamental challenges in the physics of active materials. In spite of the variety of interesting phenomena discussed in the literature, including the emergence of circulating bands \cite{Sknepnek:2015} and topological acoustic modes \cite{Shankar:2017} in polar flocks on the sphere, as well as oscillating defect patterns in spherical active nematics \cite{Keber:2014,Zhang:2016,Alaimo:2017,Khoromskaia:2017}, a coherent theoretical picture, which accounts for the threefold coupling between substrate geometry, orientational dynamics and hydrodynamic flow, is still lacking. 

In a recent work, we have investigated the dynamics of a turbulent active nematic suspension of microtubules and kinesin confined on a toroidal droplet \cite{Ellis:2018}. Using a combination of experiments and a Coulomb gas model of active defects, we demonstrated that defects in active nematics are sensitive to the Gaussian curvature of the underlying substrate, due to the passive elastic interactions between defects and curvature \cite{Bowick:2009,Turner:2010}. As turbulence progresses toward fully developed and the velocity of the defects increases, the effect of the passive elastic interactions becomes weaker and weaker, but never completely disappears. However, since the Coulomb gas model abstracts the full active nematic to a collection of point particles representing the defects, the model does not inform on how the dynamics of the active fluid itself gives rise to such effects. Without a true hydrodynamic approach, unaffected by the simplifying assumption of the particle-based models, it is impossible to achieve a complete understanding of the effect of the substrate geometry on the the spatiotemporal organization of active fluids


In this Letter we overcome this limitation and generalize the hydrodynamic theory of active nematics to arbitrarily curved substrates. By focusing on the fully developed turbulent regime, we demonstrate that topological defects can be controlled through the intrinsic geometry of the substrate.
For the specific case of the axisymmetric torus, we prove that this behavior originates from the combination of non-equilibrium effects, associated with the continuous creation/annihilation of defects, and the elastic interactions between defects and curvature. Finally, we compare our predictions with experiments on microtubule-kinesin suspensions confined to toroidal active droplets, finding excellent qualitative agreement.

\begin{figure}[b]
\includegraphics[width = \columnwidth]{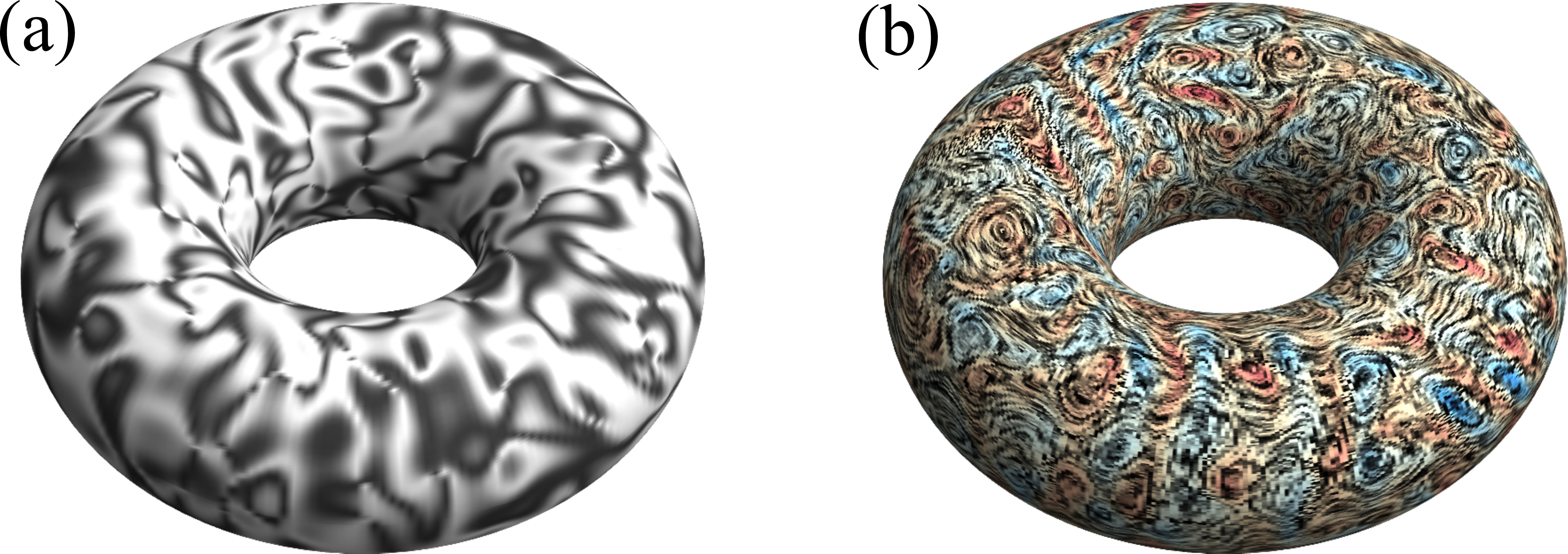}
\caption{\label{fig:snap} The (a) Schlieren texture and (b) flow on a toroidal active nematic obtained from a numerical integration of Eqs. \eqref{eq:hydrodynamics}. In (a), the polarizer and analyzer are oriented along the parallels and the meridians. Red/blue indicate areas of positive and negative vorticity.}
\end{figure}

Let $\bm{r}=\bm{r}(x^{1},x^{2})$ be the position of generic surface embedded in $\mathbb{R}^3$ and parametrized by the coordinates $(x^{1},x^{2})$. Furthermore, let $\bm{g}_{i}=\partial_{i} \bm{r}$ be a basis of covariant vectors on the surface tangent plane, so that $g_{ij}=\bm{g}_{i}\cdot\bm{g}_{j}$ is the surface metric tensor. A configuration of an active nematic liquid crystal constrained to lie on the surface can be described in terms of the local velocity field $\bm{v}=v^{i}\bm{g}_{i}$ and nematic tensor $\bm{Q}=Q^{ij}\bm{g}_{i}\bm{g}_{j}$, where $Q^{ij}=S(n^{i}n^{j}-g^{ij}/2)$, with $S$ the nematic order parameter and $\bm{n}=n^{i}\bm{g}_{i}$ the nematic director, such that $n^{i}n_{i}=1$. Incompressibility requires $\nabla_{i}v^{i}=0$, with $\nabla_{i}$ the covariant derivative. The hydrodynamic equations governing the evolution of the active nematic fluid of density $\rho$ and viscosity $\eta$ can be expressed in covariant form as follows \cite{SI}:
\begin{subequations}\label{eq:hydrodynamics}
\begin{gather}
\rho\frac{Dv^{i}}{Dt} = \eta (\Delta_{\rm B}+K)v^{i}-\zeta v^{i} + \alpha \nabla_{j}Q^{ij}\;,\\
\frac{DQ^{ij}}{Dt} = \frac{\lambda}{2}\, S u^{ij}-\frac{1}{2}\,\omega(\epsilon_{k}^{i}Q^{kj}+\epsilon_{k}^{j}Q^{ki})+\frac{1}{\gamma}\,H^{ij}\;,
\end{gather}	
\end{subequations}	
where $D/Dt=\partial_{t}+v^{k}\nabla_{k}$ is the covariant material derivative and $\Delta_{\rm B}$ is the so called Bochner or rough Laplacian. The term $\eta K v^{i}$, with $K$ the Gaussian curvature, represents the additional shear force arising from the fact that streamlines inevitably converge or diverge on surfaces with non-zero Gaussian curvature, whereas $-\zeta v^{i}$, with $\zeta$ a friction coefficient, is the damping force arising form the possible interaction of the two-dimensional active fluid with an ambient passive fluid. The last term in Eq. (\ref{eq:hydrodynamics}a) arises from the divergence of the active stress $\bm{\sigma}^{\rm a}=\alpha\bm{Q}$, where the constant $\alpha$ embodies the biochemical activity of the system. In Eq. (\ref{eq:hydrodynamics}b), $\lambda$ is the flow-alignment parameter, $u^{ij}=(\nabla^{i}v^{j}+\nabla^{j}v^{i})/2$ is the strain-rate tensor, $\omega=\epsilon_{ij}\nabla^{i}v^{j}$ is the vorticity, with $\epsilon_{ij}$ the anti-symmetric Levi-Civita tensor and $\epsilon_{i}^{j}=g^{ik}\epsilon_{jk}$, $\gamma$ is the rotational viscosity and $H^{ij}=-\delta F/\delta Q^{ij}$ is the molecular tensor describing the orientational relaxation of the system, with $F$ the free-energy. Following Kralj {\em et al.} \cite{Kralj:2011}, we express F as:
\begin{multline}\label{eq:free_energy}
F = \int {\rm d}A\,\bigg[\frac{a_{2}t}{2}\,Q_{ij}Q^{ij}+\frac{a_{4}}{4}\,(Q_{ij}Q^{ij})^{2}\;\\
+\frac{k}{2}\,\nabla_{i}Q_{jk}\nabla^{i}Q^{jk}-\frac{k_{24}}{2}\,KQ_{ij}Q^{ij}+k_{e} Q_{ij}K^{jk}K_{k}^{i}\bigg]\;,
\end{multline}
where $a_{2}$ and $a_{4}$ are constants, $t$ is the reduced temperature and is negative in the nematic phase, $K_{ij}=-\bm{g}_{i}\cdot\partial_{j}\bm{N}$, with $\bm{N}$ the normal vector, is the extrinsic curvature tensor and $k$, $k_{24}$ and $k_{e}$ are phenomenological elastic constants detailing the cost of distortions in $\bm{Q}$, the cost of having an ordered phase on a surface with Gaussian curvature, and the coupling between the order and the extrinsic curvature of the surface, respectively.

\begin{figure}[t]
\includegraphics[width = \columnwidth]{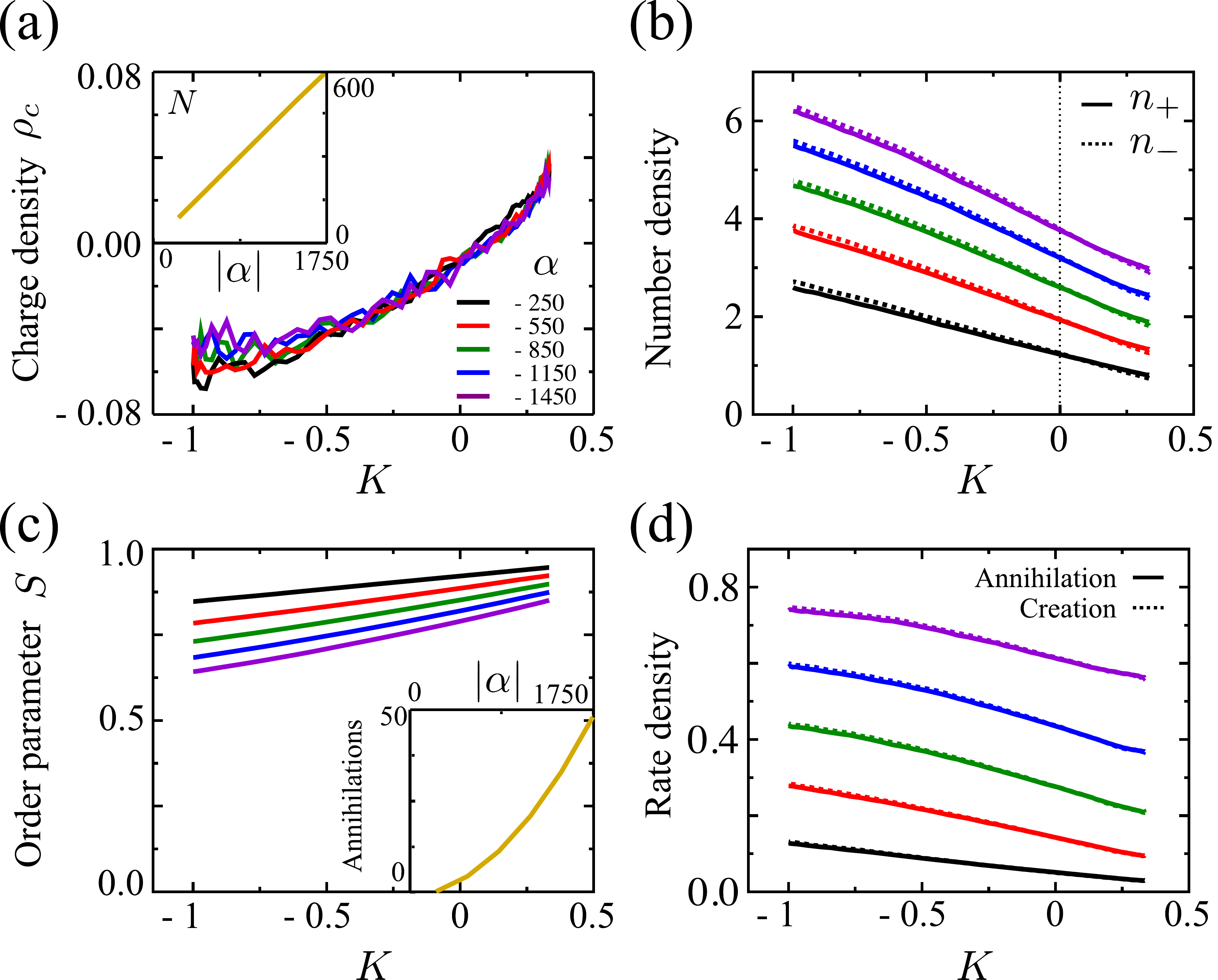}
\caption{\label{fig:HyRes} Structural properties of toroidal active nematics versus Gaussian curvature, obtained from a numerical integration of the hydrodynamics Eqs. \eqref{eq:hydrodynamics}. (a) Topological charge density $\rho_{c}$. When turbulence is fully developed, increasing the activity has little effect on $\rho_{c}$, but causes a linear increase in the number of defects (inset). (b) Number density of $+1/2$ ($n_{+}$) and $-1/2$ ($n_{-}$) disclinations. In contrast to passive liquid crystals on curved surfaces, both densities are larger in the interior of the torus, where the Gaussian curvature is negative. (c) Nematic order parameter is reduced in the interior of the torus. (d) Annihilation and creation rate densities are both increased in the interior of the torus. All quantities are rescaled as explained in the main text.}
\end{figure}

Eqs. \eqref{eq:hydrodynamics} describe the dynamics of an active nematic liquid crystal on a generic curved surface. 
Activity enters in the hydrodynamic equations only through the active force $\bm{f}^{\rm a}=\nabla\cdot\bm{\sigma}^{\rm a}$ on the right-hand side of Eq. (\ref{eq:hydrodynamics}a). Signatures of the substrate geometry are, on the other hand, imprinted in most of the terms of Eqs. \eqref{eq:hydrodynamics}, but most prominently in the molecular tensor $\bm{H}$, which affects directly the configurations of the nematic director based on how these adapt to the intrinsic and extrinsic curvature of the underlying space. We emphasize that these equations are general and well-suited to characterize any system where the individual constituents are rod-like, even if the system is not in a globally ordered phase.

To provide a concrete example and make contact with experiments, we have considered the specific case of an active nematic constrained on an axisymmetric torus (Fig. \ref{fig:snap}). Unlike the sphere, the torus is a closed surface having non-uniform Gaussian curvature. The latter is positive and a maximal on the outer equator, negative and minimal on the inner equator and varies smoothly over the surface, resulting into a vanishing Euler characteristic: $\chi = 1/(2\pi)\int K \: {\rm d}A=0$. When a torus is coated with a nematic liquid crystal, this property implies global topological charge neutrality: i.e. $\sum_{n}s_{n}=0$, with $s_{n}$ the topological charge, defined as the winding number of the nematic director along a path encircling the $n-$th defect. In practice, $s_{n}=\pm 1/2$, due to the prohibitive energetic cost of higher-charge disclinations in two dimensions. Hence the active nematic is populated by the same number of $\pm 1/2$ disclinations \cite{Sanchez:2012,DeCamp:2015,Guillamat:2016,Guillamat:2017,Ellis:2018}.

Eqs. \eqref{eq:hydrodynamics} have been numerically integrated using the vorticity/stream-function approach \cite{SI}. The integration is performed on a $256 \times 256$ square grid using finite differences and accelerated using a GPU implementation of the $V-$cycle multigrid algorithm \cite{Briggs:2000}, adapted to handle non-Euclidean metrics. To make Eqs. \eqref{eq:hydrodynamics} dimensionless, we rescale length by the cross-sectional radius of the torus, $b$, time by the relaxational time scale $k/(\gamma b^{2})$ of the $\bm{Q}-$tensor and mass by $\rho b^{2}$. In these units, we set $k = 1$, $k_{e}=0$, $k_{24}=0$, $\eta = 0.1$, $\zeta = 0.1$ and $\lambda=0.5$ in all the simulations. The aspect ratio of the tours is $\xi = a/b = 2$, with $a$ the radius of the central ring of the torus.

\begin{figure}[t]
\includegraphics[width = \columnwidth]{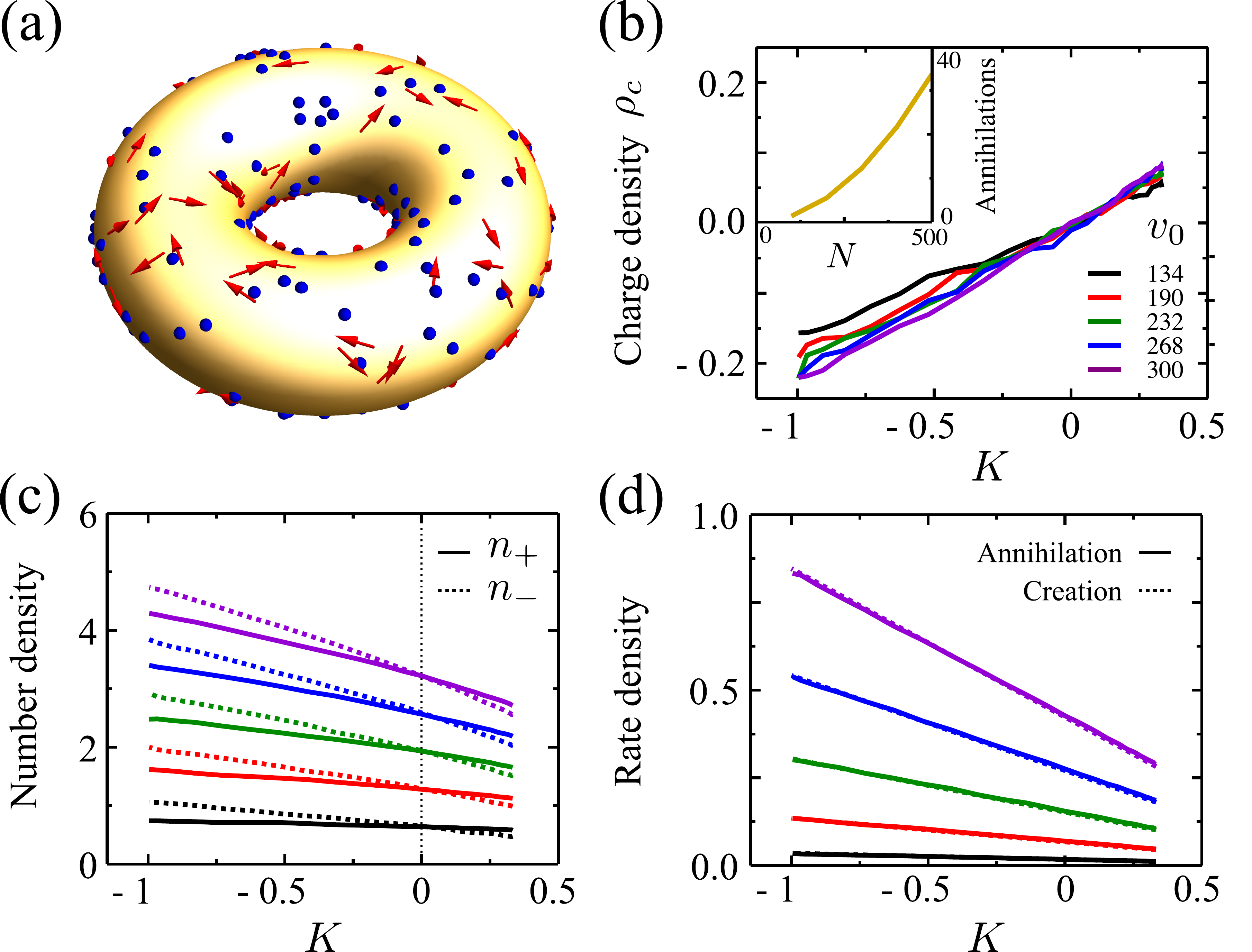}
\caption{\label{fig:PaRes} Structural properties of toroidal active nematics versus Gaussian curvature, obtained from the Coulomb gas model, Eqs. \eqref{eq:particle_model}. In all plots, length is rescaled by the cross-sectional radius $b$, time by $b^{2}/(\mu k)$ and velocity by $\mu k/b$. Activity is controlled by simultaneously varying the dimensionless defect velocity $v_{0}$ and the number $N$, taking advantage of the fact that $v_{0}\sim\sqrt{\alpha}$ and $N \sim \alpha$ (Fig. \ref{fig:HyRes}a inset). Specifically, we set $v_{0}=6\sqrt{500/N}$. (a) Snapshot of the simulation in which the topological defects are considered as point particles. (b) Topological charge density versus Gaussian curvature for varying $v_{0}$ and $N$. (Inset) The number of annihilations scales quadratically with the number of defects, hence with $\alpha$. (c) Number density of $+1/2$ ($n_{+}$) and $-1/2$ ($n_{-}$) disclinations are both linear with Gaussian curvature with negative slopes. (d) The annihilation rate density follows the same distribution as the creation rate density. 
In all simulations, $\zeta_{n}^{r}=0.1$ and $|\bm{\zeta}_{n}^{t}|=0.1$ in the units described above.}
\end{figure}

Figs. \ref{fig:snap}a,b illustrate a typical configuration of the nematic director and the vorticity. As in the case of a flat substrate, active nematics are found in a turbulent regime when the active length scale $\ell_{\rm a}=\sqrt{k/|\alpha|}$, resulting from the balance between active and passive forces, is much smaller than the system size \cite{Giomi:2015}. In this regime, the flow is organized in vortices of average size $\ell_{\rm a}$ and the director is decomposed in domains surrounded by $\pm 1/2$ disclinations. Whereas on a plane the vortices and the nematic domains are uniformly distributed in space, however, Fig. \ref{fig:snap} already suggests a correlation between the substrate geometry and the spatial organization of the coherent structures emerging within the active flow. In order to quantify this effect we have measured the time-averaged topological charge density $\rho_{c}$ of the defects, such that $\rho_{c}{\rm d}A$ is the total time-averaged topological charge in a region of area ${\rm d}A$. This is shown in Fig. \ref{fig:HyRes}a as a function of Gaussian curvature for various activity values. As hinted by Fig. \ref{fig:snap}a, the topological charge density increases monotonically with Gaussian curvature and attains its largest magnitudes along the equators. This behavior originates from the elasticity of the nematic phase, embodied in the molecular tensor $\bm{H}$ in Eq. \eqref{eq:hydrodynamics}. Assuming the nematic order parameter is constant outside the core of the defects, Eq. \eqref{eq:free_energy} approximates the one-elastic-constant Frank free-energy. In the presence of a distribution of topological defects, the latter is given by $F_{\rm F}=k/2 \int {\rm d}A\,|\nabla\varphi|^{2}$ \cite{Bowick:2009,Turner:2010}, with $\varphi$ a {\em geometric potential} given by $\Delta_{\rm LB}\varphi = \rho_{c}-K$, with $\Delta_{\rm LB}$ the Laplace-Beltrami operator \cite{SI}. This implies that the lowest energy state is attained when $\rho_{c}=K$ and $\varphi={\rm const}$. Although in the turbulent regime investigated here the system is well away from its lowest free energy state, topological defects are still subject to elastic forces attracting them to regions of like-sign Gaussian curvature, thus $\rho_{c} \sim K$. As long as active turbulence is fully developed, increasing activity has little effect on the distribution of the topological charge, but, as on the plane, increases the total number of defects (see Fig. \ref{fig:HyRes}a inset).

Whereas these observations are consistent with expectations based on the equilibrium theory of nematic order on curved surfaces, looking at the distribution of the individual $\pm 1/2$ disclinations, reveals unexpected behavior. Fig. \ref{fig:HyRes}b shows the number density $n_{\pm}$ of positive (solid line) and negative (dashed line) defects, such that $\rho_{c}=(n_{+}-n_{-})/2$. Both densities are essentially linear function of $K$ but, surprisingly, have negative slope. This implies that both the negative and positive defects are found at higher concentration in the negative Gaussian curvature region of the torus. This is in direct contrast to the passive case, in which the positive defects are only found in the positive Gaussian curvature region \cite{Bowick:2004,Jesenek:2015}. At $K=0$, $n_{+}$ and $n_{-}$ cross over, so that $\rho_{c} \sim K$, as observed in Fig. \ref{fig:HyRes}a. The nematic order parameter $S$ and the defect annihilation and creation rates also inherit the non-uniformity of the substrate geometry. Because of the higher defect concentration in the interior of the torus, $S$ is lower where $K<0$ (Fig. \ref{fig:HyRes}c). Similarly, the rate of defect creation and annihilation are higher in the interior of the torus, where the director is more distorted and defects are closer to each other (Fig. \ref{fig:HyRes}d). In addition, since the creation and annihilation rates are equivalent, the time-averaged number of defects on the torus does not change. 

We conjecture that the anomalous large number density of $+1/2$ defects where $K<0$, originates from higher distortion of the nematic director in the interior of the torus. This results into an imbalance in the defect creation rate, which is then larger in the interior. In turn, the geometrical forces due to the Gaussian curvature biases the positive (negative) topological charge toward the exterior (interior) of the torus, but, because of the short mean-free path of the defects, this does not lead to a complete segregation of the topological charge. As a consequence, the density of both positive and negative defects is larger in the interior of the torus, although their difference is proportional to $K$.

To test our conjecture, we use a variant of the Coulomb gas model of active nematic defects \cite{Keber:2014,Ellis:2018}, augmented with a non-uniform defect creation distribution reproducing the outcome of the hydrodynamic simulations. 
Defects are modelled as massless particles on the torus, whose position $\bm{r}_{n}$ and orientation $\bm{p}_{n}$ are governed by the following equations of motion:
\begin{equation}\label{eq:particle_model}
\frac{{\rm d}\bm{r}_{n}}{{\rm d}t}=v_{0}\bm{p}_{n}+\mu\bm{F}_{n}+\bm{\zeta}_{n}^{t}\;,\qquad
\frac{{\rm d}\bm{p}_{n}}{{\rm d}t}=\zeta_{n}^{r}\bm{p}_{n}^{\rm \perp}\;,	
\end{equation}
where $v_{0}$ is the speed at which defects are propelled by their self-generated flow and is zero for $-1/2$ defects and non-zero for $+1/2$ defects \cite{Giomi:2014}, $\mu$ is a mobility coefficient, $\bm{\zeta}_{n}^{t}$ and $\zeta_{n}^{r}$ are uncorrelated translational and rotational noises and $\bm{p}_{n}\cdot\bm{p}_{n}^{\perp}=0$. In addition, $\bm{F}_{n}=-\nabla_{\bm{r}_{n}}F_{\rm F}$, where $F_{\rm F}=-4\pi^{2}k\sum_{n \ne m}s_{n}s_{m}G(\bm{r}_{n},\bm{r}_{m})+2\pi k\sum_{n}s_{n}\int{\rm d}A\,G(\bm{r}_{n},\bm{r})K(\bm{r})$, with $G(\bm{r}_{n},\bm{r}_{m})$ the Laplacian Green function on the torus \cite{SI} and $\bm{F}_{n}$ is the elastic force resulting from the inter-defect interactions as well as from the interaction between the defects and the local Gaussian curvature. In the turbulent regime discussed here, $v_{0} \approx \alpha \ell_{\rm a}/\eta\sim\sqrt{\alpha}$ \cite{Giomi:2015}.

Eqs. \eqref{eq:particle_model} are solved numerically for fixed number of defects. Every time two oppositely charged defects annihilate, a new pair is created on the torus. We choose a lin- early decreasing probability distribution with the Gaus- sian curvature for pair creation, consistent with the observed trend in the hydrodyamic simulations (Fig. \ref{fig:HyRes}d). We plot $\rho_{c}$ , $n_{\pm}$ and the creation and annihilation rate densities obtained from an integration of Eqs.  \eqref{eq:particle_model} in Fig. 3. Comparing with the hydrodyamic results in Fig. \ref{fig:HyRes}, we see that the agreement is remarkable. As in our hydrodynamics simulations, the topological charge density $\rho_{c}$ from the Coulomb gas model is monotonically increasing with the Gaussian curvature (Fig. \ref{fig:PaRes}b) and essentially unaffected by the system activity (here embodied by $v_{0}$ and the number of defects $N$) in the strongly active regime discussed here. Nevertheless, the number density of both positive and negative defects, $n_{\pm}$, is higher in the interior of the torus (Fig. \ref{fig:PaRes}c), in spite of the elastic interaction between the defects and the substrate forcing the $+1/2$ defects towards the exterior of the torus. The defects annihilation and creation rates match each other exactly and are increasingly larger in the region of negative $K$ (Fig. \ref{fig:PaRes}d) as activity increases. Our particle-based simulations thus confirm our conjecture.

\begin{figure}[t]
\includegraphics[width = \columnwidth]{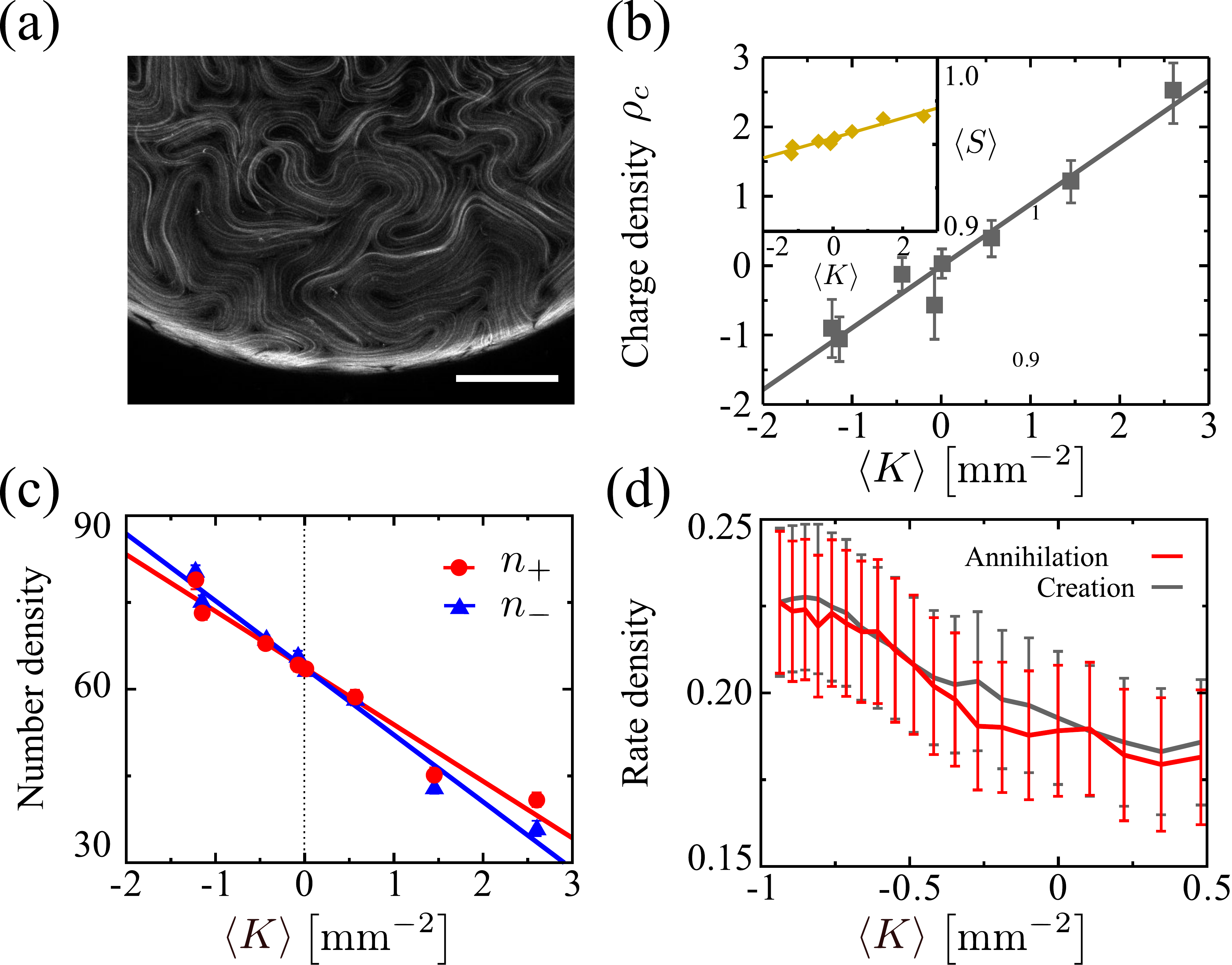}
\caption{\label{fig:ExRes} Structural properties of toroidal active nematics versus Gaussian curvature obtained from experiments with microtubules-kinesin suspensions. (a) Snapshot of the experiment. (b) Topological charge density. (c) Number density of $+1/2$ ($n_{+}$) and $-1/2$ ($n_{-}$) disclinations. (d) Annihilation and creation rates. These results correspond to observations on a torus with aspect ratio 1.8 with minor radius 334 $\mu$m. Scale bar in (a): 200 $\mu$m.}
\end{figure}

To further test the significance of our results, we compare our theoretical predictions with experiments on microtubules-kinesin suspensions constrained to the surface of toroidal droplets~\cite{Ellis:2018}. The kinesin motors are powered by adenosine triphosphate (ATP) at a concentration of $36\,\mu$M. In addition, we include an ATP regeneration system, phosphoenol pyruvate and pyruvate kinase/lactic dehydrogenase, and a depletant, polyethylene glycol (PEG), which causes the microtubules to assemble on the toroidal surface, where they form a nematic liquid crystal. We then image a portion of the lower half of the toroidal droplet using confocal microscopy and project the results along the gravitational direction onto the plane, as shown in Fig.~\ref{fig:ExRes}a. We also reconstruct both the local Gaussian curvature and $\bm{Q}$ using techniques inherited from the computer vision literature \cite{Ellis:2018}. From $\bm{Q}$ we find $S$ and $\bm{n}$, and identify defects by numerically quantifying the $\bm{n}$-rotation along a small path encircling a pixel with low $S$. We consider various regions on the surface of a given torus, calculate the mean Gaussian curvature in each region, $\langle K \rangle$, and correlate it with the time-averaged $\rho_c$ and the time-averaged $S$ in the region, as shown in Fig.~\ref{fig:ExRes}b for the example toroid in Fig.~\ref{fig:ExRes}a. We also consider the defect densities individually and correlate $n_{\pm}$ with $\langle K \rangle$ (Fig.~\ref{fig:ExRes}c). Consistent with our theoretical results (Figs. \ref{fig:HyRes} and \ref{fig:PaRes}), we find that $\rho_c$ and $S$ depend linearly on $\langle K\rangle$ with a positive slope. In addition, we also observe that $n_{\pm}$ are linearly dependent on $\langle K \rangle$ with a negative slope, corresponding to a higher topological defect density in the interior of the toroidal droplet and a lower defect density on the exterior.

At the ATP concentration used in our experiments, $\bm{Q}$ evolves slowly enough to track the defects in time using a combinatorics-based particle tracking algorithm \cite{Crocker:1996}. The individual trajectories for the $s = +1/2$ and $s = -1/2$ defects allow us to determine the creation and annihilation events; we consider the beginning and ending of a single trajectory as one-half of a defect creation or annihilation event, respectively. We then divide the number of creation and annihilation events in a region by the area of the region and the total time of the experiment to get the creation and annihilation rate density. We find that the creation and annihilation rates are equivalent, and that they are larger in regions of negative $\langle K \rangle$ than in regions with positive $\langle K \rangle$, in agreement with the theoretical results. This is shown in Fig. Fig.~\ref{fig:ExRes}d for the example toroid in Fig.~\ref{fig:ExRes}a. This agreement occurs without any dependence on the extrinsic curvature ($k_{e} = 0$) or explicit coupling between $S$ and $K$ ($k_{24} = 0$), highlighting the primary role of the intrinsic geometry.  

In summary, we have introduced a generalization of the hydrodynamic theory of active nematics to arbitrarily curved sufaces. We applied this generalization to the specific case of an extensile active nematic on the surface of a torus and probed the effect of the substrate Gaussian curvature on the active nematic. Thanks to a combination of numerical simulations and experiments we have established that the structure of the nematic phase, here described in terms of nematic order parameter, defects topological charge density, number density, creation and annihilation rates, is controlled by the substrate curvature in a two-fold way. On the one hand, defect creation and annihilation are enhanced by Gaussian curvature, leading to non-uniform nematic order and defect number density. On the other hand, the interplay between the defects and the underlying substrate geometry tends to localize the topological charge in regions of like-sign Gaussian curvature. We stress that the hydrodynamic theory in this work is general and should be able to capture phenomena in a diverse set of systems that are not necessarily deep in the nematic phase.

\begin{acknowledgments}
We would like to thank Piermarco Fonda and Gareth Alexander for helpful discussions while producing this work. This work was supported by the Netherlands Organization for Scientific Research (NWO/OCW), as part of the Frontiers of Nanoscience program and the Vidi scheme (DJGP, LG), by the National Science Foundation (NSF 1609841) and the FLAMEL program (NSF 1258425) (PWE, AFN). We thank the Brandeis biological materials facility (NSF MRSEC DMR-1420382) for providing the materials for the experimental system used in this work.
\end{acknowledgments}


\begin{thebibliography}{99}

\bibitem{Surrey:2001}	
T. Surrey, F. Nedelec, S. Leibler, E. Karsenti, 
\href{http://dx.doi.org/10.1126/science.1059758}{Science {\bf 292}, 1167 (2001)}.
	
\bibitem{Sanchez:2012}
T. Sanchez, D. N. Chen, S. J. DeCamp, M. Heymann, Z. Dogic,
\href{http://dx.doi.org/10.1038/nature11591}{Nature {\bf 491}, 431 (2012)}.
	
\bibitem{Sumino:2012}	
Y. Sumino,	K. H. Nagai, Y. Shitaka, D. Tanaka,	K. Yoshikawa, H. Chat\'e, K. Oiwa,
\href{http://dx.doi.org/10.1038/nature10874}{Nature {\bf 483}, 448 (2012)}.
	
\bibitem{Keber:2014}
F. C. Keber, E. Loiseau, T. Sanchez, S. J. DeCamp, L. Giomi, M. J. Bowick, M. C. Marchetti, Z. Dogic, and A. R. Bausch,
\href{http://dx.doi.org/10.1126/science.1254784}{\emph{Science} {\bf 345}, 1135 (2014)}.

\bibitem{Ellis:2018}
P. W. Ellis, D. J. G. Pearce, Y.-W. Chang, G. Goldsztein, L. Giomi, A. Fernandez-Nieves,
\href{https://dx.doi/org/10.1038/nphys4276}{Nat. Phys. {\bf 14}, 85 (2018).}
	
\bibitem{DeCamp:2015}
S. J. DeCamp, G. S. Redner,	A. Baskaran, M. F. Hagan, Z. Dogic,
\href{http://dx.doi/org/doi:10.1038/nmat4387}{Nat. Mater. {\bf 14}, 1110 (2015)}. 
		
\bibitem{Guillamat:2016}
P. Guillamat, J. Ign\'{e}s-Mullol, F. Sagu\'{e}s, 
\href{http://dx.doi.org/10.1073/pnas.1600339113}{Proc. Natl. Acad. Sci. U.S.A. {\bf 113}, 5498 (2016)}.

\bibitem{Guillamat:2017},
P. Guillamat, J. Ign\'{e}s-Mullol, F. Sagu\'{e}s, 
\href{http://dx.doi.org/10.1038/s41467-017-00617-1}{Nat. Commun. {\bf 8}, 564 (2017)}.

\bibitem{Schaller:2010}	
V. Schaller, C. Weber, C. Semmrich,	E. Frey, A. R. Bausch,
\href{http://dx.doi.org/10.1038/nature09312}{Nature {\bf 467}, 72 (2010)}.

\bibitem{Schaller:2013}
V. Schaller, and A. R. Bausch,
\href{http://dx.doi.org/10.1073/pnas.1215368110}{Proc. Nat. Acad. Sci. U.S.A. {\bf 110}, 4488 (2013)}.	
			
\bibitem{Dombrowski:2004}
C. Dombrowski, L. Cisneros, S. Chatkaew, R. E. Goldstein, and J. O. Kessler,
\href{http://dx.doi.org/10.1103/PhysRevLett.93.098103}{Phys. Rev. Lett. {\bf 93}, 098103 (2004)}.

\bibitem{Wogelmuth:2008}
C. W. Wogelmuth,
\href{http://dx.doi.org/10.1529/biophysj.107.118257}{Biophys. J. {\bf 95}, 1564 (2008)}.

\bibitem{Zhang:2010}
H. P. Zhang, A. Be’er, E.-L. Florin, H. L. Swinney, 
\href{http://dx.doi.org/10.1073/pnas.1001651107}{Proc. Natl. Acad. Sci. U.S.A. {\bf 107}, 13626 (2010)}.

\bibitem{Wensink:2012}
H. H. Wensink, J. Dunkel, S. Heidenreich, K. Drescher, R. E. Goldstein, H. L\"owen, and J. M. Yeomans,
\href{http://dx.doi.org/10.1073/pnas.1202032109}{Proc. Natl. Acad. Sci. U.S.A. {\bf 109}, 14308 (2012)}.

\bibitem{Dunkel:2013}
J. Dunkel, S. Heidenreich, K. Drescher, H. H. Wensink, M. B\"ar, and R. E. Goldstein,
\href{http://dx.doi.org/10.1103/PhysRevLett.110.228102}{Phys. Rev. Lett. {\bf 110}, 228102 (2013)}.

\bibitem{Wioland:2016}
H. Wioland, E. Lushi, R. E. Goldstein,
\href{http://dx.doi.org/10.1088/1367-2630/18/7/075002}{New J. Phys. {\bf 18} 075002 (2016)}.

\bibitem{Riedel:2005}
I. H. Riedel, K. Kruse, J. A. Howard, 
\href{http://dx.doi.org/10.1126/science.1110329}{Science {\bf 309}, 300 (2005)}.
		
\bibitem{Creppy:2015}
A. Creppy, O. Praud, X. Druart, P. L. Kohnke, F. Plourabou\'{e},
\href{http://dx.doi.org/10.1103/PhysRevE.92.032722}{Phys. Rev. E {\bf 92}, 032722 (2015)}.

\bibitem{Zhou:2014}
S. Zhou, A. Sokolov, O. D. Lavrentovich, and I. S. Aranson,
\href{http://dx.doi.org/10.1073/pnas.1321926111}{Proc. Nat. Acad. Sci. U.S.A. {\bf  111}, 1265 (2014)}.

\bibitem{Sknepnek:2015}
R. Sknepnek, S. Henkes,
\href{https://doi.org/10.1103/PhysRevE.91.022306}{Phys. Rev. E {\bf 91}, 022306 (2015)}.

\bibitem{Shankar:2017}
S. Shankar, M. J. Bowick, M. C. Marchetti,
\href{https://doi.org/10.1103/PhysRevX.7.031039}{Phys. Rev. X {\bf 7}, 031039 (2017)}.
 
\bibitem{Zhang:2016}
R. Zhang, Y. Zhou, M. Rahimi, J. J. de Pablo
\href{https://doi.org/10.1038/ncomms13483}{Nat. Commun. {\bf 7}, 13483 (2016)}. 
 
\bibitem{Alaimo:2017} 
F. Alaimo, C. K\"ohler, A. Voigt, 
\href{http://doi.org/10.1038/s41598-017-05612-6}{Sci. Rep. {\bf 7}, 5211 (2017)}.
   
\bibitem{Khoromskaia:2017}
D. Khoromskaia, G. P. Alexander
\href{https://doi.org/10.1088/1367-2630/aa89aa}{New J. Phys, {\bf 19}, 103043 (2017)}.   
   
\bibitem{Kralj:2011}	
S. Kralj, R. Rosso, E. G. Virga,
\href{https://doi.org/10.1039/c0sm00378f}{Soft Matter {\bf 7}, 670 (2011)}.

\bibitem{Napoli:2012a}
G. Napoli, L. Vergori, 
\href{https://doi.org/10.1103/PhysRevLett.108.207803}{Phys. Rev. Lett. {\bf 108}, 207803 (2012)}.

\bibitem{Napoli:2012b}
G. Napoli, L. Vergori, 
\href{https://doi.org/10.1103/PhysRevE.85.061701}{Phys. Rev. E {\bf 85}, 061701 (2012)}.

\bibitem{Jesenek:2014}	
D. Jesenek, S. Kralj, R. Rosso, E. G. Virga,
\href{https://doi.org/10.1039/c4sm02540g}{Soft Matter {\bf 11}, 2434 (2014)}.

\bibitem{Briggs:2000}
W. L. Briggs, V. E. Henson and S. F. McCormick, 
{\em A multigrid tutorial} 2nd ed., (SIAM, Philadelphia, 2000).

\bibitem{Giomi:2015}
L. Giomi,
\href{https://doi.org/10.1103/PhysRevX.5.031003}{Phys. Rev. X {\bf 5}, 031003 (2015)}.

\bibitem{Bowick:2009}
M. J. Bowick, L. Giomi,
\href{https://doi.org/10.1080/00018730903043166}{Adv. Phys. {\bf 58}, 449 (2009)}.

\bibitem{Turner:2010}
A. M. Turner, V. Vitelli, D. R. Nelson,
\href{https://doi.org/10.1103/RevModPhys.82.1301}{Rev. Mod. Phys. {\bf 82}, 1301 (2010)}.


\bibitem{SI}
See Supplemental Material at $\dots$

\bibitem{Bowick:2004}
M. J. Bowick, D. R. Nelson, A. Travesset,
\href{https://doi.org/10.1103/PhysRevE.69.041102}{Phys. Rev. E {\bf 69}, 041102 (2004)}.

\bibitem{Jesenek:2015}
D. Jesenek, S. Kralj, R. Rosso, E. G. Virga,
\href{https://doi.org/10.1039/C4SM02540G}{Soft Matter {\bf 11}, 2434 (2015)}.

\bibitem{Giomi:2014}
L. Giomi, M. J. Bowick, P. Mishra, R. Sknepnek, M. C. Marchetti,
\href{https://doi.org/10.1098/rsta.2013.0365}{Phil. Trans. R. Soc. A {\bf 372}, 20130365 (2014)}.

\bibitem{Crocker:1996}
J. Crocker, D. Grier,
\href{https://doi.org/10.1006/jcis.1996.0217}{J. Colloid Interface Sci. {\bf 179}, 298 (1996)}.

\end{thebibliography}
\end{document}